\documentclass[12pt]{report}
\usepackage{nmtthes2000}
\usepackage{amsfonts,amsmath,latexsym,amssymb}
\usepackage[american]{babel}

\def\II{{\mathbb I}}
\def\RR{{\mathbb R}}
\def\CC{{\mathbb C}}

\def\ZZ{{\mathbb Z}}

\def\tr{\mathrm{ tr\,}}
\def\Tr{\mathrm{ Tr\,}}

\def\Det{\mathrm{ Det\,}}

\def\vol{\mathrm{ vol\,}}
\def\R{{\mathcal R}}
\def\F{{\mathcal F}}
\def\A{{\mathcal A}}

\def\G{{\mathcal G}}
\def\L{{\mathcal L}}
\def\T{{\mathcal T}}
\def\D{{\cal D}}
\def\const{\mathrm{ const\,}}

\def\diag{\mathrm{diag}}

\def\be{\begin{equation}}
\def\ee{\end{equation}}
\def\bea{\begin{eqnarray}}
\def\eea{\end{eqnarray}}

\def\sideremark#1{\ifvmode\leavevmode\fi\vadjust{\vbox to0pt{\vss
\hbox to 0pt{\hskip\hsize\hskip1em
\vbox{\hsize2cm\tiny\raggedright\pretolerance10000
\noindent #1\hfill}\hss}\vbox to8pt{\vfil}\vss}}}

\author{Samuel Jacob Collopy}

\title{VACUUM STRUCTURE OF YANG-MILLS THEORY IN CURVED SPACETIME}
\degree{Master of Science in Physics}

\graduationdate{August, 2007}

\address{203 Garfield St. Apt. J \\ Socorro, NM 87801}

\supervisor{Ivan G. Avramidi}

\committeesize{4}

\begin{document}

    %
    %

\titlepage          

\begin{abstract}
    The stability of the chromomagnetic Savvidy vacuum in QCD  under 
    the influence of
    positive Riemannian curvature is studied. The heat traces of 
    the operators relevant to SO(2) gauge-invariant Yang-Mills fields
    and Faddeev-Popov ghosts are calculated on product spaces
    of $S^2$ and $S^1 \times S^1$. It is shown that the
    chromomagnetic vacuum with covariantly constant chromomagnetic 
    field is stable in a certain set of radii and field strengths.

    %

\end{abstract}

\begin{acknowledgments}
This thesis would not have been possible without the 
help of my thesis committee, whose collective 
tolerance of my last-minute defense allowed me to 
complete this work. I would especially like to thank
my adviser, Ivan Avramidi, who spent countless hours
explaining--and often re-explaining--every detail of 
every answer to every question that I ever posed to 
him. I owe this thesis to his efforts.
    %

\end{acknowledgments}

\tableofcontents    

\signaturepage          

\chapter{INTRODUCTION}


Quantum Chromodynamics is the highly successful 
theory of the strong interaction of elementary particles. 
It is based on $SU(3)$ Yang-Mills theory,
which is a non-Abelian gauge theory
invariant under the gauge group $SU(3)$, and a set of
spin-$1/2$ quarks, which form the fundamental representation
of the gauge group. The $SU(3)$ degrees of freedom are 
referred to as ``color.''

The high-energy behavior of Yang-Mills theory
is well-understood and leads to a renormalizable 
quantum field theory. Gross, Wilczek\cite{gross73}, and Politzer 
\cite{politzer73} discovered that at high energies (or short 
distance scales), the interaction strength of Yang-Mills fields
decreases to zero, a property known as asymptotic freedom. 

However, QCD is not well understood at low energies or large 
distance scales, i.e., those comparable to
$\Lambda_{QCD}$ ($\sim 10^{-13}$ cm), at which point
perturbation theory breaks down. The 
interaction strength increases, and the energy required to separate
quarks becomes infinite. This leads to the property of confinement,
which is exhibited by the experimental absence of free quarks, 
but has not been demonstrated 
theoretically. At energy scales less than $\Lambda_{QCD}$, 
QCD has to be replaced by an effective theory of composites 
of quarks in the form of color-neutral hadrons.

The low-energy behavior of QCD can be examined through  
the effective potential. The effective 
potential is a function of the background Yang-Mills field 
which is minimized by the absolute lowest energy state of the 
physical system. This minimum defines the physical 
vacuum.


The physical vacuum is considered trivial when the effective
potential takes a minimum with a zero background 
Yang-Mills field.
If the minimum of the effective potential 
occurs when the Yang-Mills field is non-zero, the vacuum state
will consist of a non-zero background field.
The first attempt to study the Yang-Mills vacuum in this manner
was by Savvidy\cite{savvidy77} in 1977.



Savvidy introduced a background chromomagnetic
field of constant field strength lying in the Cartan
algebra of the Lie group $SU(2)$. He found that the minimum 
of the one-loop effective 
action occurs at a non-zero value of the background field, which
causes the vacuum to be infrared unstable.

It was later pointed out by H.B. Nielsen and P. Olesen \cite{nielsen78}
that the chromomagnetic vacuum discovered by Savvidy has an
energy density with an imaginary part,
which implies that it has a tachyonic mode,
leading to instability of the vacuum. Further corrections\cite{nielsen79}
were made to the chromomagnetic vacuum to show that the energy
is lower when the chromomagnetic vacuum consists of tube-like domain
structures, with the chromomagnetic field pointed along the axis
of each tube. The finite width of the tubes serves as an infrared
cutoff, which destroys the low-energy instability. The minimum 
energy density of this type of state has been found by 
\cite{nielsen79b} to be a superposition of domains separated by
a fixed distance.
This model is known as the ``spaghetti vacuum.''

 
It is important to note that these calculations pertain to a 
chromomagnetic field in flat space.
In this paper, we will consider the related problem of the
stability of the Yang-Mills vacuum on a curved space.
The addition of the curvature will make the operator corresponding to
the second variation of the action
positive definite for large enough values of curvature,
which will change the tachyonic mode into a physical state and cause
the vacuum to stabilize. 
In particular, we will consider non-zero covariantly constant
$SO(3)$ Yang-Mills fields on product spaces of spheres,
valued in the sub-algebra $SO(2)$. By 
analyzing the spectrum of the second variation of the action and
computing the effective action, it will be shown
that the vacuum will stabilize on spaces that have sufficiently strong
curvature.

This paper is organized as follows.
Chapter Two of this paper discusses the effective action 
approach to quantum field theory and how it applies to gauge 
theories, as well as the technique of zeta-function regularization.
In order to compute the effective action, we need the spectrum 
of the Yang-Mills and Faddeev-Popov ghost operators on spheres, 
which are calculated in Chapter Three.  
In Chapter Four, we apply the results of
Chapter Three to find the heat trace of Yang-Mills and ghost
operators on $S^2$ and $S^1 \times S^1$. Chapter Five is 
devoted to finding the total heat kernel on the products of
spheres and determining in what cases the vacuum 
is stable.



\chapter{QUANTIZATION OF NON-ABELIAN GAUGE THEORIES}



In this chapter, we will use the proper time method of
Schwinger and DeWitt, and so the notation will follow that
of DeWitt\cite{dewitt65}.

\section{Kinematics}

\subsubsection{Spacetime Geometry}


The spacetime under consideration is an $n$-dimensional
pseudo-Rie\-man\-nian  manifold $M$ endowed with a
globally hyperbolic metric $g$ with signature
$\left(- + \dots +\right)$.
We will assume that the spacetime manifold has a global
time-like Killing vector so that $M = \RR \times \Sigma$,
where $\Sigma$ is an $(n-1)$-dimensional compact oriented spin
manifold without boundary. Local coordinates $x^{\mu}$ on $M$
are labeled by Greek indices that run over $0,\dots,n-1$. The
coordinate basis $\partial_{\mu}$ for the tangent space $T_xM$
at the point $x \in M$ has dual basis $dx^{\mu}$ in $T_x^*M$.

The Christoffel symbols can be found from the metric 
\be
\Gamma^\alpha{}_{\beta\gamma} = \frac{1}{2}g^{\alpha\delta}\left(\partial_\gamma g_{\delta\beta} + \partial_\gamma g_{\delta\gamma} - \partial_\delta g_{\beta\gamma}\right)\,.
\ee
The curvature of the metric $g_{\mu\nu}$ is described by the Riemann curvature tensor
\be
R^{\alpha}{}_{\mu\beta\nu} = \partial_{\beta}\Gamma^{\alpha}{}_{\nu\mu}
- \partial_{\nu}\Gamma^{\alpha}{}_{\beta\mu}
+ \Gamma^{\eta}{}_{\nu\mu}\Gamma^{\alpha}{}_{\beta\eta}
- \Gamma^{\eta}{}_{\beta\mu}\Gamma^{\alpha}{}_{\nu\eta}\,,
\ee
and its contractions: the Ricci tensor,
\be
R_{\mu\nu} = R^{\alpha}_{\phantom{\alpha}\mu\alpha\nu}\,,
\ee
and the Ricci scalar,
\be
R = g^{\mu\nu}R_{\mu\nu}\,.
\ee

\subsubsection{Orthonormal Frame}
An orthonormal frame $e_{(\alpha)}$ can be constructed at every point
on the manifold and is labeled by lower case Greek indices 
in parentheses.
The orthonormal frame
of $T_xM$ can be constructed as a set of vector fields 
over $M$, where $\alpha = 1,\dots,n$ so that
\be
\left<e_{(\alpha)},e_{(\beta)}\right> = \eta_{(\alpha)(\beta)}\,,
\ee
where $\eta_{(\alpha)(\beta)} = \diag\left(-1,1,\dots,1\right)$.

The orthonormal basis $e_{(\alpha)}$ may be expanded in the coordinate
basis $\partial_{\mu}$,
\be
e_{(\alpha)} = e_{(\alpha)}{}^{\mu}\partial_{\mu}\,.
\ee
The inverse matrix $e^{(\alpha)}{}_{\mu}$ of $e_{(\alpha)}{}^{\mu}$ defines the dual basis
\be
e^{(\alpha)} = e^{(\alpha)}{}_{\mu}dx^{\mu}
\ee
in the cotangent space $T^*_xM$. Then
\be
g^{\mu\nu}e^{(\alpha)}{}_{\mu} e^{(\beta)}{}_{\nu} = \eta^{(\alpha)(\beta)}\,,
\qquad
g_{\mu\nu}e_{(\alpha)}{}^{\mu}e_{(\beta)}{}^{\nu}=\eta_{(\alpha)(\beta)}\,.
\ee

\subsection{Gauge Group}

Yang-Mills theory describes the dynamics of a vector bundle
over $M$. To say that the theory is gauge invariant is to
impose the restriction that the action does not
change under transformations by a gauge group.
In particular, consider a compact simple Lie group
$G$ attached to every point of $M$ so that any neighborhood in a
fiber bundle has the local structure $M \times G$.
In other words
let $k^a$ be coordinates on $G$, so that for any coordinate patch
on $M$, a point in the fiber bundle is described by the
set of coordinates $\left(x^{\mu},k^a\right)$. Group
indices are labeled by lower case Latin letters, which run over
$1,\dots,\dim G$.

An element $U \in G$ of a compact simple gauge group may be written in the form
\be
U = \exp{\left(k^a T_a\right)}\,,
\ee
where $k^a$ are parameters and $T_a$ are the generators of the
Lie group $G$, which lie in the Lie algebra. It is clear that the
expression
\be
T_a = \frac{d}{dk^a}U|_{k^a = 0}
\ee
is an equivalent definition of the generators of $G$.
The generators
of a simple compact Lie algebra satisfy the relation
\be
[T_a,T_b] = C^c_{\phantom{c}ab}T_c\,,
\ee
where $C^c_{\phantom{c}ab}$ are the structure constants of the Lie
algebra $G$.

The adjoint representation of the Lie algebra is defined by taking
the generators to be
\be
\left(T_a\right)^b{}_c =C^b{}_{ac}\,.
\ee

To form inner products between algebra-valued tensors,
we must introduce an inner product on the Lie algebra. We define
the Cartan-Killing metric
\be
E_{ab} = -\frac{1}{2}C^c_{\phantom{c}ad}C^d_{\phantom{d}bc}
 = -\frac{1}{2}\tr(T_a T_b)
\ee
to raise and lower group indices. In the case of compact simple
Lie groups, this metric can be normalized by
\be
E_{ab} = \delta_{ad}\,.
\ee

To make the action invariant under gauge
transformations, the covariant derivative $\nabla_{\mu}\varphi^{\nu}$
of a field must satisfy the condition
\be
\nabla'_{\mu}\varphi'^\nu = U (\nabla_\mu\varphi^\nu) \,.
\ee
The gauge matrix $U$ is local, i.e., depends on the coordinates,
and primed quantities denote the transformed quantities.
To do this, we let
\be
\nabla_\mu\varphi^\nu = (\nabla^{LC}_{\mu} + \A_{\mu})\varphi^\nu\,,
\ee
where $\A_\mu = A^a_\mu T_a$ is an algebra-valued vector field that transforms as
\be
\A'_\mu = U \A_\mu U^{-1} -\left(\partial_{\mu}U\right) U^{-1}\,
\ee
for any gauge transformation $U \in G$.
In general, $\A_\mu$ depends on the representation of the group $G$.

The strength $\F_{\mu\nu} = F^a_{\mu\nu} T_a$ of the Yang-Mills field $\A_\mu$ is defined by
\be
\F_{\mu\nu} = \partial_\mu\ A_\nu - \partial_\nu \A{}_\mu + [\A_\mu,\A_\nu]\,.
\ee
The strength of the field can be used to define the action functional
\be
S_{YM} = \frac{1}{8e^2}\int_{M} dx\,\tr(\F^{\mu\nu}\F_{\mu\nu})\,,
\ee
where $e$ is a coupling constant and $\tr$ denotes the trace over
the Lie algebra.

\subsection{Scalar Fields}

A scalar field $\varphi$ is invariant under diffeomorphisms and
has a covariant derivative of
\be
\nabla_\mu\varphi = (\partial_\mu + \A_\mu)\varphi\,,
\ee
with $\A$ in an appropriate representation. 
The action for a scalar field must be constructed out of
a scalar potential term $V(\varphi)$ 
and the quantity $\varphi^T\Box\varphi$, where T denotes
transpose and $\Box$ is the D'Alembert operator
\be
\Box \equiv \nabla^\mu \nabla_\mu\,.
\ee

\section{Effective Action}
The effective action approach to quantum field theory is a highly
useful approach that was developed by DeWitt and others
\cite{dewitt65,dewitt03,vilkovisky_gospel,avramidi00}. This section follows
the method as developed for boson fields.

Consider two causally connected in- and out- regions of spacetime
that lie in the past and future of a region $\Omega$ in which
physical dynamics will take place. The goal of quantum field
theory is to compute the amplitude $\left<{\rm in}|{\rm out}\right>$
for some initial state $\left|{\rm in}\right>$ in the in- region to evolve
into some final state $\left|{\rm out}\right>$ in the out region. To
calculate this, consider a change in the action $\delta S$. The
Schwinger variational principle states that the amplitude
$\left<{\rm in}|{\rm out}\right>$ will change according to
\be
\delta \left<{\rm in}|{\rm out}\right>
= \frac{i}{\hbar}\left<{\rm in}|\delta S|{\rm out}\right>\,.
\ee

Let $\varphi^i$ be the boson fields relevant to the problem, where
$i$ is taken to run over both continuous (i.e. spacetime) and discrete
(i.e. spinor, tensor, field) indices. Change the action by adding a
linear interaction of $\varphi^i$ with some classical sources $J_i$
that vanish in the in- and out- regions
$\delta S = J_i\varphi^i$, where the contraction over $i$ is taken as
both a summation over discrete indices and integration
over spacetime
\be
\varphi^i J_i = \int_M dx\;\sqrt{g}\;
\varphi_{(A)}J^{(A)}\,,
\ee
where $g = \det g_{\mu\nu}$.

With this variation, the solution to the Schwinger variational
principle is expressed in terms of the Feynman path integral
\be
\label{pathint}
\langle {\rm out} | {\rm in} \rangle= \int {\cal D}\varphi
\exp \left\{\frac{i}{\hbar}
[S(\varphi)+J^k\varphi_k]\right\}\,,
\ee
where ${\mathcal D}\varphi$ represents the Feynman measure.

The generating functional for connected diagrams
$W(J)$ is defined in terms of the in-out transition amplitude by
\be
\label{defW}
\left<{\rm out}|{\rm in}\right> \equiv \exp\left(\frac{i}{\hbar}W(J)\right)\,.
\ee
The first functional derivative of $W$ gives the background field $\Phi^i$
\be
\Phi^i(J) = \frac{\delta}{\delta J_i} W(J)\,,
\ee
the second functional derivative produces the propagator
\be
\G^{i_1 i_2}(J) = \frac{\delta^2}{\delta J_{i_1} \delta J_{i_2}}W(J)\,,
\ee
and the higher derivatives produce the many-point Green functions
\be
\G^{i_1 \dots i_k}(J) = \frac{\delta^k}{\delta J_{i_1} \dots \delta J_{i_k}}W(J)\,.
\ee

In order to calculate vertex functions, we define the effective action
$\Gamma(\Phi)$ by the functional Legendre transform
\be
\label{effac}
\Gamma(\Phi) = W(J(\Phi)) - J_i(\Phi)\Phi^i\,,
\ee
where the sources $J(\Phi)$ are expressed in terms of the
background fields. The first functional derivative of $\Gamma$ is
equal to the sources
\be
\frac{\delta}{\delta \Phi^i} \Gamma(\Phi) = -J_i(\Phi)\,,
\ee
the second derivative defines the inverse propagator
\be
\frac{\delta^2}{\delta \Phi^{i} \delta \Phi^{j}}
\Gamma(\Phi) = D_{ij}(\Phi)\,,
\ee
\be
D_{ij}\G^{jk} = -\delta_i{}^k\delta(x,x')\,,
\ee
and the higher derivatives determine the vertex functions
\be
\Gamma_{i_1 \dots i_k}(\Phi) \equiv
\frac{\delta^k}{\delta \Phi_{i_1}\cdots \delta \Phi_{i_k}}\Gamma(\Phi)\,.
\ee

We see from (\ref{pathint}), (\ref{defW}), and (\ref{effac}) that the
effective action satisfies
\be
\exp\left\{\frac{i}{\hbar}\Gamma(\Phi)\right\} =
\int {\cal D}\varphi \exp\left\{\frac{i}{\hbar}\left[S(\varphi)
-\frac{\delta\Gamma(\Phi)}{\delta\Phi^i}(\varphi^i - \Phi^i)\right]\right\}\,.
\ee

\subsection{Gauge Theory}
In gauge field theories, the above formalism can not be applied
immediately. Instead, there will be problems arising from the fact
that the measure will include an integral over non-physical fields.
To be more explicit, consider an action functional
$S(\varphi)$
that is invariant under some
vector fields
$\mathbf{R}_{A} = R^i{}_A(\varphi)\delta/\delta\varphi^i$
on the configuration space. Transformations of the fields
\be
\delta_{\xi}\varphi^i = R^i{}_A\xi^A\,,
\ee
with $\xi^A$ being some parameters,
that do not affect any real physics
are called gauge transformations.

In the case of gauge theories, the Feynman path integral (\ref{pathint})
will be carried over both physical and non-physical
degrees of freedom. This adds divergences to the path integral that
can only be removed by the DeWitt-Fadeev-Popov method.
We can separate the field variables $\varphi$ into physical
variables $I^{(A)}$ and gauge variables $\chi^{B}$,
so that the action $\bar S(I)=S(\varphi(I,\chi))$
does not depend on the group variables $\chi$, that is,
it is invariant under the variations with respect to $\chi^{B}$,
but not under the variations with respect to $I^{(A)}$.

To get rid of the excess degrees of freedom, we change the variables
$\varphi=\varphi(I,\chi)$ in the path integral and omit the integration
over the group variables $\chi$ (in other words, we divide out the
volume of the gauge group), so that the integral is over the
physical variables $I^{(A)}(\varphi)$ only.
If the inverse change of variables is given by $\chi=\chi(\varphi)$,
then the Jacobian of the change of variables is $\Det Q(\varphi)$, where
\be
Q^A{}_B(\varphi) = R^i{}_{B}(\varphi)\frac{\delta\chi^A(\varphi)}{\delta \varphi^i}
\ee
is the Faddeev-Popov operator.
The gauge condition is defined by a surface in the configuration
space
\be
\chi^A(\varphi)=\theta^A
\ee
where $\theta^{B}$ are some constants.
This surface intersects all
orbits of the gauge group transversally, and thus has a one-to-one
correspondence with the set of all physical states.
With these
changes, we get the Feynman measure
\be
\D I = \D \varphi \Det Q(\varphi)
\delta(\chi^{B}(\varphi) - \theta^{B})\,,
\ee
where $\delta(\chi^{B} - \theta^{B})$ is the functional delta function.
The path integral becomes
\bea
\exp\left(\frac{i}{\hbar}\Gamma(\Phi)\right) &=&
\int {\cal D}\varphi\,\delta(\chi^{B}(\varphi) - \theta^{B})\Det Q(\varphi)
\nonumber
\\[5pt]
&&
\times \exp \left\{\frac{i}{\hbar}\left[S(\varphi) -\frac{\delta\Gamma(\Phi)}{\delta\Phi^i}
(\varphi^i - \Phi^i) \right]\right\}\,.
\eea
In order to write the effective action in terms of gauge-invariant
quantities, we integrate over $\theta^{B}$ with Gaussian weight to get
the expression
\bea
&&
\exp\left(\frac{i}{\hbar}\Gamma(\Phi)\right) =
\int {\cal D} \varphi \Det Q(\varphi)(\Det H(\Phi))^{\frac{1}{2}}
\nonumber
\\[5pt]
&&
\label{effacc_gauge}
\qquad
\times \exp \left\{\frac{i}{\hbar}\left[S(\varphi)
- \frac{1}{2}\chi_{A}(\varphi)H^{AB}(\Phi)\chi_{B}(\varphi) -\frac{\delta\Gamma(\Phi)}{\delta\Phi^i}
(\varphi^i - \Phi^i) \right]\right\}\,,
\nonumber\\
\eea
where $H^{AB}$ is some non-degenerate operator that does not
depend on $\varphi$. The determinant $\Det Q$ is
typically calculated in terms of Faddeev-Popov ghost fields, and $\Det H$ 
is calculated in terms of the Nielsen-Kallosh ghost.
In the case that $H$ does not depend on the background field $\Phi$,
$\det H$ is simply an infinite constant that can be factored
into the measure.

\section{Perturbation Theory}

In order to calculate the effective action, we introduce perturbation
theory. In this section, we follow \cite{avramidi02}. Perturbation
theory is based on the idea that the largest contributions to the
effective action come from fields $\varphi^i$ close to the background
field $\Phi^i$ in the sense that they can be split into a background part and
a quantum part
\be
\label{phisplit}
\varphi^i = \Phi^i + \sqrt{\hbar}\;h^i\,,
\ee
with $h^i$ being the quantum fluctuations of the background field.
With this change of variables, the effective action will have an
expansion of the form
\be
\label{gammaseries}
\Gamma(\Phi) = S_{YM} + \hbar\Gamma_{(1)} + O(\hbar^2)\,,
\ee
which is known as the loop expansion.

Splitting the fields as in (\ref{phisplit}), the path integral
(\ref{effacc_gauge}) may be expanded using (\ref{gammaseries}) to 
give the first order correction
\be
\label{gamma1gauge}
\Gamma_{(1)} = \frac{i}{2}\ln\Det L_{YM} -\frac{i}{2}\ln\Det H -i\ln\Det L_{FP}\,,
\ee
where
\be
\label{LYM_secvar}
(L_{YM})_{ik} = -S_{,ik}(\Phi) +
\chi_{A}{}_{\,,i}(\Phi)H^{AB}\chi_{B}{}_{\,,k}(\Phi)\,,
\ee
and
\be
L_{FP} = Q(\Phi)\,.
\ee
The quantity $\Gamma_{(1)}$ is the one-loop effective action.
%

%
\subsubsection{Wick rotation}

The determinants of differential operators $\ln\Det L$ 
are problematic for two reasons. One reason is
that $\ln\Det L$ is divergent and must be regularized in order to make
sense. This will be dealt with in the next section. Before that
can be accomplished we must first deal with the fact that $L$
is, in general, not a positive definite operator. To get rid of
this latter problem, we Wick rotate time in the complex plane
\be
t = -i\tau\,.
\ee

Under this transformation, the metric is changed to have the
Riemannian signature $(+,\dots,+)$, so the operator $-\Box$
becomes an elliptic operator rather than a hyperbolic one.
In addition, the measure picks up an additional factor of $-i$:
\be
\sqrt{|g|}\,{\rm d}t\,{\rm d}^{n-1}x \to -i\sqrt{|g|}\,{\rm d}\tau\,{\rm d}^{n-1}x\,,
\ee
which in turn causes the action to pick up the same factor.
For the remainder of the paper, all quantities will be assumed
to be Wick rotated.
To describe finite temperature effects, the Euclidean time is 
compactified to $S^1$, with radius given by the inverse 
temperature $\beta = 1/T$.

\section[Heat Kernel Approach]{Heat Kernel Method for
Computing the One-Loop Effective Action}

The quantity $\Gamma_{(1)}$ in (\ref{gamma1gauge})
is the functional determinant of an elliptic differential operator. This
quantity will always be infinite and therefore must be regularized
in order for any physical calculation to make sense.
Following \cite{vassilevich03,avramidi00}, this section
will show that using the heat kernel representation, the effective
action can be expressed in terms of a zeta function. 
Then by analytic continuation, the zeta function can be
regularized and made to yield finite physical results.
\subsubsection{Green Functions}

For a bosonic field, the second functional derivative of
the action may be brought by choice of gauge to the form
\be
\label{minimal}
L + m^2= -\Box + Q +m^2\II\,,
\ee
where $Q$ is a matrix-valued function acting on the fields $\varphi^i$,
$m$ is the mass (which may be zero), and $\II$ is the identity matrix.

Green functions are solutions $\G^j{}_k$ of the equation
\be
(L +m^2)\G(x,x') = \II\delta(x,x')\,,
\ee
with
\be
\delta(x,x') = g^{-1/2}(x)\delta(x-x')\,.
\ee
They can be constructed in terms of a contour integral of the heat kernel
$U(t) \equiv U(t|x,x')$
\be
\label{greenheat}
\G(x,x') = \int_0^\infty {\rm d}t\exp(-tm^2)U(t|x,x')\,.
\ee
The heat kernel satisfies the heat equation
\be
\label{heateq}
\left(\frac{\partial}{\partial t} + L\right)U(t) = 0\,
\ee
with the boundary condition
\be
\label{heatbound}
\left.U(t|x,x')\right|_{t=0} = \II\delta(x,x')\,.
\ee

\subsubsection{From Green functions to Effective Action}
The heat equation (\ref{heateq}) has the formal solution
\be
U(t) = \exp\left(-tL\right)\,.
\ee
In terms of the eigenfunctions $\phi_n$ corresponding
to eigenvalues $\lambda_n$ of the operator $L$, i.e.
\be
L\phi_n = \lambda_n\phi_n\,,
\ee
the heat kernel can be written
\be
U(t|x,x') = \sum^\infty_{n=1}\phi_n(x)\otimes\phi^{\dagger}_n(x')e^{-t\lambda_n}\,.
\ee
The heat kernel diagonal is defined by taking the coincidence limit
of this expression
\be
U(t|x,x) = \sum^\infty_{n=1}\phi_n(x)\phi_n^\dagger(x)e^{-t\lambda_n}\,,
\ee
and the functional $L^2$ heat trace is the trace of the diagonal over all indices
\bea
\Tr \exp(-tL) &=& \int_M {\rm d}x\, \tr U(t|x,x)
\nonumber
\\
\label{heattrace}
&=& \sum^\infty_{n=1}e^{-t\lambda_n}\,,
\eea
In the case of Yang-Mills theory,
the trace $\tr$ is over both group indices and tangent space indices.
If the eigenvalues are degenerate, we may express the heat kernel 
in terms of the eigenvalues $\{\lambda_n\}$and degeneracies $\{d_n\}$
\be
\label{heat_trace}
\Tr \exp(-tL) = \sum^\infty_{n=1}d_n e^{-t\lambda_n}
\ee

If the mass is sufficiently large so that $\lambda_n + m^2 > 0$,
the quantity $\ln \Det(L +m^2)$ can similarly be expressed
\be
\ln \Det[L +m^2] =
\sum^\infty_{n=0}\ln(\lambda_n+m^2)\,.
\ee
We can use the identity
\be
\ln \lambda = -\int^{\infty}_{0}\frac{dt}{t}e^{-t\lambda} + C
\ee
with $C$ an infinite constant,
and the expression for the heat trace (\ref{heattrace}) to find
\be
\label{lndet}
\ln \Det (L + m^2) =
-\int_0^\infty\frac{dt}{t}\exp(-tm^2)\int_M dx\, \tr U(t|x,x) + \const \,.
\ee
The infinite constant has no effect on the dynamics, and can be dropped.
The one-loop effective action is expressed completely in terms
of the logarithms of determinants of operators, so calculation of the heat kernel $U(t|x,x)$ for various
operators gives all of the information needed to calculate
$\Gamma_{(1)}$. This reduces the task of calculating the one-loop
effective action to that of finding the eigenvalues of the 
second variation of the action.

\subsubsection{Zeta-Function Regularization}
The quantity $\ln\Det (L + m^2)$ (\ref{lndet}) is infinite.
In order to make it finite, it must be regularized
in terms of the $\zeta-$function
\be
\label{lndetzeta}
\ln\Det (L + m^2) = -\zeta'(0)\,,
\ee
\be
\zeta'(0) = \frac{\rm d}{{\rm d}p}\zeta(p)|_{p=0}\,.
\ee

The $\zeta-$function of a differential operator $M$ is defined in terms
of the heat kernel by
\be
\zeta_M(p) = \mu^{2p}\Tr M^{-p} = \frac{\mu^{2p}}{\Gamma(p)}
\int_0^{\infty}dt\, t^{p-1}\Tr \exp(-tM)\,,
\ee
where $\mu$ is a renormalization parameter with dimension of 
inverse length.
The $\zeta-$function is analytic at $p=0$, so the expression
(\ref{lndetzeta}) is finite and well-defined.

\subsection{Yang-Mills One-Loop Effective Action}
With this method of regularization in mind, we may return
to Yang-Mills theory.
The one-loop effective action for Yang-Mills theory (\ref{gamma1gauge})
in a general covariant gauge is the sum of contributions
from Yang-Mills fields and ghosts.
In terms of functional determinants in Euclidean space,
\be
\label{gamma1ym}
\Gamma_{(1)} = \ln\Det L_{YM}(\lambda)
- 2\ln\Det L_{FP}(\lambda)\,,
\ee
where $\lambda$ is a gauge-fixing parameter,
and $L_{YM}(\lambda)$ is the operator acting on gauge fields
as found in (\ref{LYM_secvar})
\be
L_{YM}(\lambda) = L_{YM} + \lambda H\,,
\ee
\be
(L_{YM}\varphi)^{\mu} = -\Box\varphi^{\mu} - 2\F^{\mu}{}_{\nu}\varphi^\nu + R^{\mu}{}_{\nu}\varphi^\nu
\label{LYM}
\ee
\be
(H\varphi)^{\mu} = \nabla^{\mu}\nabla_{\nu}\varphi^\nu\,,
\ee
$L_{FP}$ is the Faddeev-Popov ghost operator acting on anti-commuting scalar fields
\be
L_{FP}(\lambda) = \sqrt{1-\lambda}L_{FP},
\ee
\be
\label{LFP}
L_{FP}\eta = -\Box\eta\,.
\ee

We can regularize the gauge-fixed $\Gamma_{(1)}$ by
expressing it in terms of the $\zeta-$function
\be
\label{gamma1zeta}
\Gamma_{(1)} = -\frac{1}{2}\zeta'_{\rm tot}(0)\,,
\ee
where
\be
\label{zeta_total}
\zeta_{\rm tot}(p) = \zeta_{L_{YM}}(p) -2\zeta_{L_{FP}}(p)
\ee
is the total $\zeta-$function.
The zeta-function can be analytically continued to give a 
renormalized expression for the effective action.

The factor
$\sqrt{1-\lambda}$ guarantees gauge independence of the
regularized effective action on the mass shell\cite{avramidi95a}.
It can also be
proven \cite{avramidi95a} that $\zeta_{\rm tot}$ is independent
of gauge, and so we may choose $\lambda = 0$ so that we are left
with minimal differential operators as in (\ref{minimal}).

%
%
%

\section{Chromomagnetic Vacuum}

It was shown by Savvidy that the one-loop effective action for
$SU(2)$ Yang-Mills in flat space takes a minimum for a non-zero field.
Consider a covariantly constant Yang-Mills background in flat
space
\be
\nabla_\mu\F_{\alpha\beta} = 0\,.
\ee
One flat space solution to this equation is
\be
\A^a{}_\alpha = -\frac{1}{2}F_{\alpha\beta}x^\beta\,n^a,
\ee
so that $F^a_{\mu\nu}$ takes the form $F^a{}_{\alpha\beta} = F_{\alpha\beta}n^a$,
where $n_b$ is a unit vector in the Cartan subalgebra of the Lie
algebra of $G$, $n^bn_b = 1$.
To make this a ``magnetic'' background, conditions on group invariants
are imposed
\be
F_{\alpha\beta}F^{\alpha\beta} = \frac{1}{2}(H^2 - E^2) > 0\,,
\ee
\be
\epsilon^{\alpha\beta\gamma\delta}F_{\alpha\beta}F_{\gamma\delta}
= H\cdot E = 0\,.
\ee
Expanding the one-loop effective action 
in terms of momenta
for a magnetic-type field and taking the first term\cite{savvidy77},
the one-loop correction is
$\Gamma_{(1)H} = \int dx \L_{(1)H}$
\be
\L_{(1)H} =
\frac{1}{8\pi^2e^2} \int_0^\infty \frac{{\rm d}s}{s^3} \left(\frac{Hs}{\sinh(Hs)}
+ 2Hs\,\sin(Hs)\right)\,.
\ee
Renormalization of this expression gives
\be
\L_{(1)H} = -\frac{11H^2}{48\pi^2}
\left[\ln\frac{H}{\mu^2} - \frac{1}{2}\right]
\ee
to which the corresponding energy density is
\be
{\mathcal H} = \frac{H^2}{2e^2} + \frac{11H^2}{48\pi^2}
\left[\ln\frac{H}{\mu^2} - \frac{1}{2}\right]\,.
\ee
It is easily seen \cite{nielsen78} that the energy density has
a minimum at
\be
H_{\rm min} = \mu^2\exp \left(-\frac{24\pi^2}{11e^2}\right)\,.
\ee

It was later pointed out by Nielsen and Olesen \cite{nielsen78}
that the energy density for this model has an imaginary part,
which leads to instability of this model.

%
\chapter{HEAT TRACE ON SPHERES}

\section{Existence of covariantly constant Yang-Mills fields on spheres}
In the spirit of Savvidy \cite{savvidy77}, we consider a
covariantly constant field strength tensor $\F_{\mu\nu}$
that takes values in the center of the gauge Lie algebra.
The condition that the field is constant gives rise to the
equation
\be
[\nabla_\mu,\nabla_\nu]\F_{\alpha\beta} -
[\nabla_\alpha,\nabla_\beta]\F_{\mu\nu} = 0\,.
\ee
It can be shown\cite{avramidi94} that this yields the
integrability condition
\be
\label{integrable}
[\F_{\mu\nu},\F_{\alpha\beta}] +
R_{\mu\nu\lambda[\alpha}\F^\lambda{}_{\beta]}
- R_{\alpha\beta\lambda[\mu}\F^\lambda{}_{\nu]} = 0\,.
\ee
By taking $\F_{\mu\nu}$ to be in the center of the Cartan algebra,
$[\F_{\mu\nu},\F_{\alpha\beta}] = 0$, writing the Riemann tensor
of the $N-$sphere as
\be
R^{\mu\nu}{}_{\lambda\alpha} = \rho(\delta^\mu_\lambda\delta^\nu_\alpha
    -\delta^\nu_\lambda\delta^\mu_\alpha)
\ee
for $N \ge 2$,
and contracting over $\mu$ and $\alpha$ in (\ref{integrable}),
we find
\be
\rho(N-2)F_{\nu\beta} = 0\,.
\ee
Therefore, non-zero covariantly constant magnetic fields in the center of
the algebra of the Lie group can only exist on $S^2$ or $\RR^N$.

\section{Product Manifolds}
Each of the manifolds that we consider has the product
manifold structure $M = M_1 \times \dots \times M_n$,
where $M_i$ are submanifolds of $M$.
In this case, we have the decomposition of the operator $L$
\be
L = L_1 \otimes \II_2 \otimes \dots \otimes \II_n + \dots +
\II_1 \otimes \dots \II_{n-1} \otimes L_n\,,
\ee
where $L_i$ is the projection of $L$ onto the submanifold $M_i$.
In this case, the heat kernel has the form
\be
\exp(-tL) = \exp(-tL_1) \dots \exp(-tL_n)\,.
\ee
To calculate the heat kernel on a general product manifold,
we only need to calculate the heat kernel on each submanifold
and multiply the results
\be
\Tr e^{-tL} = \Tr_{M_1}\exp(-tL_1) \dots \Tr_{M_n}\exp(-tL_n)\,,
\ee
where $\Tr$ denotes the functional trace, which is also
taken to be the trace over group and coordinate indices.

\section{Heat Trace of Laplacian on $S^1$}
On $S^1$, a circle of radius $r$,
the Laplacian acting on any function is 
simply the operator
\be
L = -\Delta = -\frac{1}{r^2}\partial^2_\phi\,,
\ee
where $\phi$ is the coordinate along the circle, 
$0 \le \phi < 2\pi$.
The eigenvalues for this operator are
\be
\lambda_n = \frac{n^2}{r^2}\,,\qquad n = 0, \pm 1, \dots
\ee
The multiplicities are 
\bea
&&
d_0 = 1\\
&&
d_n = 2\,, \qquad n = 1,2,\dots
\\
\eea
The heat kernel trace for a function on $S^1$ can then
be calculated using the formula for the heat kernel (\ref{heat_trace})
\be
\Tr \exp(-tL) = 1 + 2\sum_{n=1}^\infty e^{-tn^2/r^2}\,.
\ee
We define this to be the function
\be
S\left(\frac{t}{r^2}\right) = 1 + 2\sum_{n=1}^\infty e^{-tn^2/r^2}\,.
\ee
There is no difference between a scalar and a $p$-form on
$S^1$, so this heat trace applies to all geometric objects
on $S^1$. 
\section{Heat Trace on $S^2$}
The heat trace for $S^2$ is non-trivial and its calculation is
a significantly more complicated problem. On $S^2$, scalars and 
one-forms will be distinct objects and will form different
representations of both the gauge group and the isotropy group,
which will determine the covariant derivative and thus the form
of the Laplacian. In addition, the existence of non-zero 
chromomagnetic fields will cause the the eigenvalues to split, 
leading to a much more complicated spectrum.

\subsection{Geometry of $S^2$}
Consider the 2-sphere $S^2$ of radius $R$
endowed with the standard Riemannian metric
\be
ds^2 = e^{(\alpha)}{}_\mu e^{(\alpha)}{}_\nu dx^\mu dx^\nu
     = R^2(d\theta^2 + \sin^2\theta d\phi^2)\,,
\ee
where $0 \le \theta < \pi$, and $0 \le \phi < 2\pi$.
Greek letters without parentheses denote the coordinate indices, which range
over the two values $\theta$ and $\phi$.
The orthonormal basis one-forms $e^{(\alpha)}$ are given by
\be
e^{(1)} = R d\theta\,,\qquad e^{(2)} = R \sin\theta d\phi\,.
\ee
Greek indices with parentheses
range over $1,2$ and denote indices of the orthonormal basis.
The volume form is given by
\be
d{\rm vol} = R^2\sin\theta d\theta \wedge d\phi\,.
\ee
The components of the spin connection 1-form can be
found by the Cartan method to be
\be
\omega_{(\alpha)(\beta)} = - \epsilon_{(\alpha)(\beta)}\cos\theta d\phi\,.
\ee
The curvature tensor components are
\be
R^{(\alpha)(\beta)}{}_{(\gamma)(\delta)} = \frac{1}{R^2}
    \left(\delta^{(\alpha)}_{(\gamma)} \delta^{(\beta)}_{(\delta)}
    - \delta^{(\beta)}_{(\gamma)} \delta^{(\alpha)}_{(\delta)}\right)\,.
\ee

\subsection{Isometries}

The sphere $S^2$ is 
diffeomorphic to the quotient space $SO(3)/SO(2)$. 
Here, $SO(3)$ is the isometry group of $S^2$ and $SO(2)$ is the 
isotropy group of $S^2$. The rotation group $SO(3)$
maps $S^2$ to itself and $SO(2)$ rotations centered around a point
will leave the point unmoved.

Let $\varphi = (\varphi^A)$ be a field which transforms
under a representation of the group $SO(2)$.
Let $\Sigma_{(\alpha)(\beta)}
    = \left(\Sigma_{(\alpha)(\beta)}{}^B{}_A\right)$
be the generators of the group $SO(2)$
in the representation acting on $\varphi$.
Because $SO(2)$ is a one-dimensional, Abelian group,
there is only one generator $\Sigma = \Sigma_{(1)(2)}$.
Then the covariant derivative of $\varphi$ is
\be
\nabla_\mu\varphi =
    \left(\partial_\mu + \omega_\mu\Sigma\right)\varphi\,,
\ee
where
\be
\omega_1 = 0\qquad \omega_2 = -\cos\theta\,.
\ee

\subsection{Gauge Curvature}
Now assume that the field $\varphi$ also transforms under
another representation $T$ of the group $SO(2)$, which we call a 
gauge representation. Then $\varphi$ transforms under the
product of two representations of the group $SO(2)$.
Let $\A$ be the corresponding gauge connection and
$\F = d\A$ be the curvature of this connection.

The gauge curvature $\F$ is a 2-form on a 2-dimensional
space, and so $\F$ must be proportional to the volume form
$d{\rm vol}$
\be
\F = \frac{TH}{2}\sin\theta d\phi \wedge d\theta\,,
\ee
where, in general, $H$ is some function of the coordinates.
By expressing $\F$ in components
\be
\label{f_coords}
\F^\mu{}_\nu = \frac{TH}{2R^2}E^\mu{}_\nu \,,
\ee
where $E_{\mu\nu} = \sqrt{g}\epsilon_{\mu\nu}$ is the
invariant volume form, we see
\be
\nabla_{\alpha} \F_{\mu\nu} = \frac{1}{2}
T E_{\mu\nu}\nabla_{\alpha}H\,.
\ee
Requiring the curvature to be covariantly
constant yields the condition that $H$ is a constant.

Physically, $H$ can be interpreted as the charge of
a monopole at the center of the sphere.
Because $\F$ is precisely
the Chern form of a line bundle over $S^2$ 
\cite{frankel97}, we have
\be
\int\limits_{S^2}\F = 2\pi nT\,,\quad n \in \ZZ\,.
\ee
Therefore, the monopole charge $H$ can only take
integer values
\be
H = n \,,\quad n \in \ZZ\,.
\ee

The corresponding gauge connection is found by
solving the equation
\be
\F = d\A\,,
\ee
which results in
\be
\A = \frac{TH}{2}\cos \theta d\phi
\ee
We see that $\A$ is proportional to the
spin connection $\omega$.
The covariant derivative invariant under both space
rotations and gauge transformations is then
\be
\nabla_\mu\varphi = (\partial_\mu + \T \omega_\mu)\varphi\,,
\ee
where
\be
\label{generator}
\T = \II\otimes\Sigma + \frac{TH}{2}\otimes\II\,,
\ee
and $T$ is the generator of the gauge group $SO(2)$.

\section{Spectrum of the Laplacian on $S^2$}
In order to calculate the heat traces (\ref{heat_trace})
on $S^2$, we need to analyze the spectrum of the operator $L_{YM}$.
We know from (\ref{LYM}) that this operator is equal to
the negative Laplacian plus Yang-Mills strength and Ricci curvature
terms. The field strength and curvature tensors are covariantly constant.
Therefore, the eigenfunctions of $L_{YM}$ are
proportional to the eigenfunctions of the Laplacian. This causes the
eigenvalues of $L_{YM}$ to be the eigenvalues of the Laplacian, shifted 
by the eigenvalues of the sum of the other two operators.

The Laplacian acting on general spin-tensor is
given by the expression
\be
\label{Laplacian}
\Delta = g^{\mu\nu}\nabla_\mu\nabla_\nu
    = |g|^{-1/2}(\partial_\mu + \T\omega_\mu)
    |g|^{1/2}g^{\mu\nu}(\partial_\nu + \T\omega_\nu)\,.
\ee
In the case of $S^2$, the Laplacian becomes
\be
\Delta = \frac{1}{R^2}\left[\partial_\theta^2
+ \cot\theta \partial_\theta
+ \frac{1}{\sin^2\theta}(\partial_\phi - \T\cos\theta)^2\right]
\ee

It can be noted that this will yield the standard Laplacian
for a particle in a magnetic field \cite{landau77} in the
limit $R \to \infty$.
Using polar coordinates
near $\theta = 0$ with $\theta = \rho/R$,
along with using the choice of gauge
$\A = \frac{tH}{2}\left(\cos\theta - 1\right)d\phi$ and denoting the generator
of $SO(2)$ by $i$,  
the connection becomes
\be
\A_{(1)} = 0\,,\qquad \A_{(2)} = \frac{iMR^2}{2}(\cos\theta - 1)\,,
\ee
where $M$ is the magnetic field $H = MR^2$.
Taking the limit $R \to \infty$ then gives the 
standard Laplacian on $\RR^2$
\be
\label{l_low}
\Delta = \partial_\rho^2 + \frac{1}{\rho}\partial_\rho + \frac{1}{\rho^2}\partial_\phi^2 
+ \frac{iH}{2}\partial_\phi - \frac{H^2}{16}\rho^2\,.
\ee

\subsection{Action of Laplacian on one-forms}

For 1-forms, the generator
$\T$ is the matrix with components
\be
\T^{(\alpha) a}{}_{(\beta) b} = \epsilon^{(\alpha)}{}_{(\beta)}\delta^{a}{}_{b}
    + \frac{H}{2}\epsilon^{a}{}_{b}\delta^{(\alpha)}{}_{(\beta)}\,.
\ee
The eigenvalues of the matrix $\T$ are $i k_j\,, j =1,2,3,4$
\be
k_1 = 1 + \frac{H}{2}\,,\quad
k_2 = 1 - \frac{H}{2}\,,\quad
k_3 = -1 + \frac{H}{2}\,,\quad
k_4 = -1 - \frac{H}{2}\,.
\ee
In the same basis, the field strength tensor can be
diagonalized with corresponding eigenvalues
\be
f_1 = -\frac{H}{2}\,,\quad
f_2 = \frac{H}{2}\,,\quad
f_3 = \frac{H}{2}\,,\quad
f_4 = -\frac{H}{2}\,,
\ee
and the Ricci tensor will be proportional to the identity,
with all eigenvalues $r_i$ given by
\be
r_i = \frac{1}{R^2}
\ee
Diagonalizing the matrix $\T$ will cause the Yang-Mills operator
$L_{YM}$ to break into four separate operators of the form
\be
L^{(i)} = -\frac{1}{R^2}\left[\partial_\theta^2
+ \cot\theta \partial_\theta
+ \frac{1}{\sin^2\theta}(\partial_\phi - i k_i\cos\theta)^2\right]
-2\frac{f_i}{R^2} + \frac{1}{R^2}\,,
\ee
where $k$ is a half-integer that takes one of the
four values $k_1,k_2,k_3,k_4$. The values $f_i$ are the 
corresponding eigenvalues of the matrix $\F$ and are 
given by the values
\be
f_1 = -\frac{H}{2}\,,\qquad
f_2 = \frac{H}{2}\,,\qquad
f_3 = \frac{H}{2}\,,\qquad
f_4 = -\frac{H}{2}\,.
\ee
It is clear that the spectrum of the Laplacian will be 
invariant under change of sign of the field $H$.
For the remainder of the paper, we will then assume 
without loss of generality that $H$ is positive.

The spectrum of the operator $L^{(i)}$ is
defined by regular normalized solutions of the equation
\be
L^{(i)}u = \lambda u\,,
\ee
where $\lambda\in \CC$ is a complex spectral parameter.
This equation has regular solutions only for certain real
discrete values of $\lambda$, which 
determine the spectrum of $L^{(i)}$.

The operators $\F$ and $R$ have no dependence on the coordinates,
and so the eigenfunctions of the operator $L^{(i)}$ are the same 
as for the Laplacian. The eigenvalues of $L^{(i)}$ are obtained 
from the eigenvalues of the Laplacian $(-\Delta)$ by shifting 
\be
\label{shift}
\lambda(L^{(i)}) = \lambda(-\Delta^{(i)}) + \frac{1}{R^2} -\frac{2f_i}{R^2}\,,
\ee
so we find the spectrum of the Laplacian first.

Separating variables with the substitution
\be
u = e^{i m \phi}h_m(\theta)\,,\quad m \in \ZZ\,,
\ee
we obtain an ordinary differential equation for $h(\theta)$
\be
\left\{\partial_\theta^2
+ \cot\theta \partial_\theta
-\frac{1}{\sin^2\theta}
( m -  k_i\cos\theta)^2 + R^2\lambda \right\}h_m(\theta) = 0\,.
\ee
Let us introduce the notation
\be
a^\pm_{ml} = \frac{1}{2} + \left|\frac{m-k}{2}\right|
    + \left|\frac{m+k}{2}\right| \pm \frac{1}{2}(1 + 4R^2\lambda)^{1/2}\,.
\ee
The index $l$ labels the eigenvalues, as will be described below.
As explained in Appendix A, this equation has regular normalized
solutions given by
\bea
h^l_{mk}(\theta) &=&
    (1-\cos\theta)^{\left|\frac{m-k}{2}\right|}
    (1+\cos\theta)^{\left|\frac{m+k}{2}\right|}
\nonumber
\\
&&
\times F\left(a^{+}_{ml},a^{-}_{ml}; 1+\left|m-k\right|;\, \frac{1-\cos\theta}{2}\right)\,,
\nonumber
\\&&
\eea
where $F(a,b;c;z)$ is the hypergeometric function.
In the case of integer $k$, these solutions exist for
the following values of $\lambda_l$ and $m$
\be
\label{lambda_int}
\lambda_{l} = \frac{1}{R^2}\,(|k| + l)(|k|+l+1)\,,
\ee
\be
-l \le m \le l\,,
\ee
where $l$ is an integer greater than or equal to $0$:
\be
l \ge 0\,.
\ee
By counting all possible values of $m$ we obtain the multiplicities
of the eigenvalues $\lambda_l$ for integer $k$
\be
\label{deg_int}
d_l = 2(l + |k|)+1\,.
\ee
In the case of half-integer $k$,
there are two series of solutions. The first series is given by
the following values of $\lambda_l$ and $m$:
\bea
&&
\label{lambda_half1}
\lambda_l = \frac{1}{R^2}\,(|k| + l)(|k|+l+1)\,,
\\
&&
-|k| + \frac{1}{2} \le m \le |k|-\frac{1}{2}\,,
\eea
where $l$ is an integer greater than or equal to $0$
\be
l \ge 0\,,
\ee
giving degeneracies
\be
\label{deg_half1}
d_l = 2|k|\,.
\ee
The second series is given by
\bea
&&
\label{lambda_half2}
\lambda_n = \frac{1}{R^2}\,\left(|k|
+ \frac{1}{2} + n\right)\left(|k|+\frac{3}{2}+n\right)\,,
\\[10pt]
&&
-\left(|k| + \frac{1}{2}+n\right) \le m
\le -\left(|k| + \frac{1}{2}\right)
\mbox{ or }
\nonumber\\[10pt]
&&
\left(|k| + \frac{1}{2}\right) \le m
\le \left(|k| + \frac{1}{2} + n\right)\,,
\eea
which gives the degeneracies
\be
\label{deg_half2}
d_n = 2n+2\,,\quad n \ge 0\,.
\ee
Therefore, the eigenvalues of the operator $(-\Delta_j)$ are $\lambda_l$
and the corresponding eigenfunctions are
\be
u^l_{m}(\theta,\phi) = e^{im\phi} h^l_{m}(\theta)\,.
\ee

It should be noted here that the eigenvalues of the Laplacian
here are very different from those of the Laplacian for flat 
space (\ref{l_low}). In the flat space, the eigenvalues are
that of a harmonic oscillator\cite{landau77}
\be
\lambda = HR^{-2}\left(n + \frac{1}{2}\right)\,,
\ee
whereas in our case, the eigenvalues are quadratic in $|k|$,
and so will increase quadratically in $H$ as $H$ becomes large.

The spectrum of the operator $L_{YM}$ is then obtained by the shift
of the Laplacian's eigenvalues (\ref{shift}).
The heat kernel of the operator $L_{YM}$ acting on one-forms 
is found using the eigenvalues of the Laplacian, which are
the values of $\lambda$ in (\ref{lambda_int})-
(\ref{lambda_half2}) plus the value $1/R^2 - 2f_i/R^2$.

The eigenvalues for even magnetic field are given by 
\be
\lambda_{il} = \frac{1}{R^2}(|k_i|+l)(|k_i|+l+1) + \frac{1}{R^2} - \frac{2f_i}{R^2}\,,
\ee
where $i$ runs over the tangent space and group indices $i = 1,2,3,4$,
$l\ge0$, and the degeneracies are given by
\be
d_{il} = 2(l+|k_i|) + 1\,.
\ee
The eigenvalues for odd magnetic field are in two series for each 
value of $i$. The first series is given by
\be
\lambda_{il} = \frac{1}{R^2}(|k_i| + l)(|k_i| + l + 1) + \frac{1}{R^2} - \frac{2f_i}{R^2}
\ee
with $i=1,2,3,4$, $l\ge0$, and degeneracies given by
\be
d_{il} = 2|k_i|\,.
\ee
The second series is given by
\be
\lambda_{in} = \frac{1}{R^2}\left(|k_i| + \frac{1}{2} + n\right)\left(|k_i| + \frac{3}{2} + n\right)\,,
\ee
with $i=1,2,3,4$, $n\ge0$, and degeneracies
\be
d_{in} = 2n+2\,.
\ee


\subsection{Action of Laplacian on Ghosts}
In addition to the Yang-Mills field, we must consider
the scalar Faddeev-Popov ghost field to compute the
effective action. 
In this case, the relevant operator is $L_{FP} = -\Delta$.
Scalar fields are invariant under coordinate transformations,
so the generator $\T$ as given in (\ref{generator}) 
will only transform under the gauge group SO(2), which
means that the total generator for the ghosts will be
\be
X = \frac{H}{2}\epsilon^a{}_b\,,
\ee
which has the two eigenvalues $i\kappa_j$ with
\be
\label{ghost_k}
\kappa_1 = \frac{H}{2}\,,\qquad \kappa_2 = -\frac{H}{2}\,.
\ee
The values that the parameter $\lambda$ takes on will 
be exactly the same as in
(\ref{lambda_int}), (\ref{lambda_half1}),
and (\ref{lambda_half2}), except with these values of $\kappa$ to 
replace the values of $k$. The eigenvalues are again characteristically
different for even and odd magnetic field.
For even values of $H$, the eigenvalues of $L_{FP}$ are given by
\be
\lambda_l = \frac{1}{R^2}\left(l + \frac{H}{2}\right)\left(l + \frac{H}{2} + 1\right)
\ee
The eigenvalues do not depend on the sign of $\kappa$, so the 
degeneracies are just doubled to account for the two values:
\be
d_l = 4\left(l + \frac{H}{2}\right) + 2\,.
\ee
For odd values of $H$, the eigenvalues are in two series.
The first series is given by
\be
\lambda_l = \frac{1}{R^2}\left(l + \frac{H}{2}\right)\left(l + \frac{H}{2} + 1\right)\,,
\ee
with the degeneracies doubled to account for positive and negative $kappa$:
\be
d_l = 2H\,.
\ee
The second series is given by
\be
\lambda_n = \frac{1}{R^2}\left(n + \frac{H}{2} + \frac{1}{2}\right)\left(n + \frac{H}{2} + \frac{3}{2}\right)\,,
\ee
with degeneracies
\be
d_n = 4n + 4\,.
\ee
By using these values, we will be able to calculate the heat kernel
in the next chapter.

%
\chapter{HEAT KERNEL TRACE AND EFFECTIVE ACTION}
The heat trace of the total Yang-Mills and ghost operators 
can be computed on products of spheres using the factorization
property of the heat kernel. In this chapter, we calculate
the heat traces of each operator on the spaces $T^2$ and $S^2$,
from which we will be able to compute the total heat trace
for the product manifolds in the next chapter.
The heat kernel can readily be 
calculated from the
eigenvalues and degeneracies that have been found,
as well as using the defining formula for the heat trace
(\ref{heattrace}), written in terms of eigenvalues $\lambda_l(L)$ and
their degeneracies,
\be
\label{sum}
\Tr(e^{-tL}) = \sum_l d_l e^{-t\lambda_l(L)}\,.
\ee

\section{Yang-Mills on $T^2$}
The simplest space to consider is the two-torus, 
$T^2 = S^1 \times S^1$,
with each copy of $S^1$ having a different radius 
$(r_1, r_2)$. The
two-torus has no curvature and can support no 
covariantly constant chromomagnetic field due to
topological constraints. 
Thus, the operator $L_{YM}$ is just the Laplacian
\be
L_{YM} = -\Delta\,.
\ee
It is then straightforward to find that the heat trace is
\be
\label{T2_YM}
\Tr \exp(-tL_{YM}) = 4S\left(\frac{t}{r_1^2}\right)S\left(\frac{t}{r_2^2}\right)\,.
\ee

The operator for ghosts in this case is simply the scalar Laplacian $-\Delta$,
\be
\label{T2_FP}
\Tr \exp(-tL_{FP}) = 2S\left(\frac{t}{r_1}\right)S\left(\frac{t}{r_2}\right)\,,
\ee
with the factor of $2$ coming from the trace over group indices.
%
%


\section{Yang-Mills on $\RR^2$}
The heat trace for Yang-Mills theory on $\RR^2$ has been found 
\cite{avramidi93,avramidi94} to be 
\be
\label{R2_YM}
\Tr e^{-tL_{YM}} = \int_{\RR^2} dx (4\pi t)^{-1} \left[2 + \frac{tHR^{-2}}{\sinh(tHR^{-2}/2)}
\left(2 + 4\sinh^2(tHR^{-2}/2)\right)\right]
\ee
and the heat trace for the corresponding ghost operator is 
\be
\label{R2_FP}
\Tr e^{-tL_{FP}} = \int_{\RR^2} dx (4\pi t)^{-1} \left[1 + \frac{tHR^{-2}}{\sinh(tHR^{-2}/2)}\right]\,.
\ee

\section{Yang-Mills on $S^2$}

\subsection{Yang-Mills Operator on $S^2$}
The heat kernel for the Yang-Mills field can be found by
performing the spectral sum (\ref{sum}), using the eigenvalues
and degeneracies for the operator $L_{YM}$.

For the case of even magnetic charge $H$, the heat kernel for the
operator $L_{YM}$ acting on one-forms is
\bea
Tr(e^{-tL_{YM}}) &=& \sum_{j=1}^4 \sum_{l=0}^\infty (2l + 2|k_j| +1)
\nonumber\\
&&
    \times\exp\left\{-\frac{t}{R^2}\left[
    (|k_j| + l)(|k_j|+l+1) -2f_j + 1\right]\right\}\,.
\eea
For odd $H$,
\bea
&&
Tr(e^{-tL_{YM}}) = \sum_{j=1}^4 \Bigg[
\sum_{l=0}^\infty 2|k_j|
\exp\left\{-\frac{t}{R^2}\left[(|k_j|+l)(|k_j|+l+1) -2f_j +1\right]\right\}
\nonumber\\[10pt]
&&
\qquad
+ \sum_{l=0}^\infty (2l+2)\exp\left(-\frac{t}{R^2}
\left(|k_j|+l + \frac{1}{2}\right)\left(|k_j|+l+\frac{3}{2}\right)
-2f_i + \frac{1}{R^2}\right)
\Bigg]
\nonumber\\
&&
\eea
These sums can be expressed in terms of the functions
\be
\Theta_j(t) = \sum^\infty_{l=1} l^j e^{-tl(l+1)}\,,
\ee
\be
\Phi_j(t) = \sum^\infty_{l=1} l^j e^{-tl^2}\,.
\ee
These functions are regular in the limit $t \to \infty$.\\
For $H = 0$, the heat trace is given by
\be
\label{S2_H0}
\Tr(e^{-tL_{YM}}) = 4e^{-t/R^2}\left[2\Theta_1\left(\frac{t}{R^2}\right)
    + \Theta_0\left(\frac{t}{R^2}\right)\right]\,.
\ee
For $H = 1$,
\bea
\Tr(e^{-tL_{YM}}) &=& (4e^{-2t/R^2} + 4)\Theta_1\left(\frac{t}{R^2}\right)
-4e^{-2t/R^2}\Theta_0\left(\frac{t}{R^2}\right)
\nonumber\\
&&
+(6e^{-9t/4R^2} + 2e^{-t/4R^2})\Phi_0\left(\frac{t}{R^2}\right)
-6e^{-13t/4R^2}
\eea
For $H = 2$,
\be
\label{S2_H2}
\Tr(e^{-tL_{YM}}) = 2(e^{t/R^2} + e^{-3t/R^2})\left[2\Theta_1\left(\frac{t}{R^2}\right)
    + \Theta_0\left(\frac{t}{R^2}\right)\right]
    + 2e^{t/R^2} - 6e^{-5t/R^2}\,.
\ee
For $H = 3$,
\bea
\Tr(e^{-tL_{YM}}) &=& 
(2e^{7t/4R^2} + 10e^{-19t/4R^2})\Phi_0\left(\frac{t}{R^2}\right)
+(4e^{2t/R^2} +4e^{-4t/R^2})\Theta_1\left(\frac{t}{R^2}\right)
\nonumber\\
&&
-10e^{-23t/4R^2} - 4e^{-6t/R^2} - 10e^{-35t/4R^2} - 4e^{-10t/R^2}\,.
\label{S2_H3}
\eea
For $|H| = 4$,
\be
\Tr(e^{-tL_{YM}}) = 2(e^{3t/R^2} + e^{-5t/R^2})
 \left[2\Theta_1\left(\frac{t}{R^2}\right) + \Theta_0\left(\frac{t}{R^2}\right)\right]
 -6e^{-7t/R^2} - 10e^{-11t/R^2}\,.
\ee
For $|H| \ge 5$ with $|H|$ odd,
\bea
&&
\Tr(e^{-tL_{YM}}) = 8\cosh(tH/R^2)e^{-t/R^2}\Theta_1\left(\frac{t}{R^2}\right)
\nonumber\\
&&
\qquad
+ [(6-2H)e^{t(H-1)/R^2} - 2(H+1)e^{-t(H+1)/R^2}]\Theta_0\left(\frac{t}{R^2}\right)
\nonumber\\
&&
\qquad
+ [(2H+4)e^{t(H-3/4)/R^2} + (2H+4)e^{-t(H+3/4)} - 8e^{t(H-5/4)/R^2}]\Phi_0\left(\frac{t}{R^2}\right)
\nonumber\\
&&
\qquad
-2e^{-t(H+3/4)/R^2}(H+2)\sum_{l=1}^{\frac{H}{2}+\frac{1}{2}}e^{-tl^2/R^2}
\nonumber\\
&&
\qquad
-2e^{t(H-3/4)/R^2}(H-2)\sum_{l=1}^{\frac{H}{2}-\frac{3}{2}}e^{-tl^2/R^2}
\nonumber\\
&&
\qquad
-2e^{-t(H+1)/R^2}\sum_{l=1}^{\frac{H}{2}+\frac{1}{2}}(2l-H-1)e^{-tl(l+1)/R^2}
\nonumber\\
&&
\qquad
-2e^{t(H-1)/R^2}\sum_{l=1}^{\frac{H}{2}-\frac{3}{2}}(2l+3-H)e^{-tl(l+1)/R^2}
\label{S2_H5}        
\eea
For $|H| \ge 6$ with $|H|$ even,
\bea
\Tr(e^{-tL_{YM}}) &=& 4\cosh(tH/R^2)e^{-t/R^2}\left[2\Theta_1\left(\frac{t}{R^2}\right)
	-\Theta_0\left(\frac{t}{R^2}\right)\right]
\nonumber\\
&&
    -2e^{-t(H+1)/R^2}\sum_{l=1}^{\frac{H}{2}}(2l+1)e^{-tl(l+1)/R^2}
\nonumber\\
&&
    -2e^{t(H-1)/R^2}\sum_{l=1}^{\frac{H}{2}-2}(2l+1)e^{-tl(l+1)/R^2}\,.
\label{S2_H6}
\eea

%
%
%
%

\subsection{Ghost Operator on $S^2$}
In addition to the Yang-Mills field itself, there are ghost
fields to eliminate the extra degrees of freedom caused by
gauge invariance.
The ghost fields on $S^2$ are scalar fields, which means that 
the eigenvalues and degeneracies are given by 
(\ref{lambda_int})-(\ref{deg_half2}) with the values 
$k = \pm |H|/2$. With these values, it is straightforward
to calculate the heat trace.\\
For H = 0,
\be
\label{S2_H0g}
\Tr (e^{-tL_{FP}}) = 4\Theta_1\left(\frac{t}{R^2}\right)
	+ 2\Theta_0\left(\frac{t}{R^2}\right) + 2\,.
\ee
For $H = 1$,
\be
\label{S2_H1g}
\Tr (e^{-tL_{FP}}) = 2e^{-t/4R^2}\Phi_0\left(\frac{t}{R^2}\right) + 4\Theta_1\left(\frac{t}{R^2}\right)\,.
\ee
For $H = 2$,
\be
\label{S2_H2g}
\Tr (e^{-tL_{FP}}) = 4\Theta_1\left(\frac{t}{R^2}\right) + 2\Theta_0\left(\frac{t}{R^2}\right)\,.
\ee
For $H$ odd, with $|H| \ge 3$,
\bea
\Tr (e^{-tL_{FP}}) &=& 2He^{t/4R^2}\Phi_0\left(\frac{t}{R^2}\right) 
+ 4\Theta_1\left(\frac{t}{R^2}\right)
+ 2(1-H)\Theta_0\left(\frac{t}{R^2}\right)
\nonumber\\
&& 
-2He^{t/4R^2}\sum_{l=1}^{\frac{H}{2}-\frac{1}{2}}e^{-tl^2/R^2}
-\sum_{l=1}^{\frac{H}{2}-\frac{1}{2}}(4l+2-2H)e^{-tl(l+1)/R^2}
\nonumber\\
&&
\label{S2_H3g}
\eea
For even $H$, $|H| \ge 4$, we have
\be
\label{S2_H4g}
\Tr e^{t\Delta_0} = 4\Theta_1\left(\frac{t}{R^2}\right)
+ 2\Theta_0\left(\frac{t}{R^2}\right)
-2\sum_{l=1}^{\frac{|H|}{2}-1}(2l+1)e^{-\frac{t}{R^2}l(l+1)}\,.
\ee
\section{Heat Trace on Product Spaces}
With the heat trace of both the Yang-Mills and ghost operators
calculated on all relevant submanifolds, it is possible to 
assemble the total heat trace the gauge-fixed Yang-Mills
field on four-dimensional manifolds by using the factorization
property of the heat kernel and calculating the total heat trace
$\Tr \exp(-tL_{YM}) - 2\Tr \exp(-tL_{FP})$\,.

We can characterize the stability of the Yang-Mills vacuum by 
examining the large $t$ behavior if this function. A negative
eigenvalue corresponds to an unstable mode, which would
indicate that the vacuum is unstable. However, we find that
with a sufficiently strong positive curvature on $S^2$, we can 
make all eigenvalues positive, and the vacuum becomes stable.

\section{Yang-Mills on $S^1 \times S^1 \times \RR^2$ with 
non-zero chromomagnetic field on $\RR^2$}
This case is the analog of the problem studied by Savvidy. 
Spacetime has zero curvature and a constant chromomagnetic field
exists.
The total heat kernel is given by
\bea
U_{\rm tot}(t) &=& \Tr \exp(-tL_{YM}){}_{\RR^2}\times\Tr \exp(-tL_{YM}){}_{S^1 \times S^1}
\nonumber\\
&&
-2\,\Tr \exp(-tL_{FP}){}_{\RR^2}\times\Tr \exp(-tL_{FP}){}_{S^1 \times S^1}\,.
\nonumber\\
&&
\eea
This is evaluated using the heat traces (\ref{R2_YM}) and (\ref{R2_FP}):
\be
U_{\rm tot}(t) = \int_{\RR^2} dx (4\pi t)^{-1} \left[\frac{tHR^{-2}}{\sinh(tHR^{-2}/2)}
\left(4\sinh^2(tHR^{-2}/2)\right)\right]
S\left(\frac{t}{r_1^2}\right)\,S\left(\frac{t}{r_2^2}\right)\,,
\ee
where $r_1$ and $r_2$ are the radii of the two copies of $S^1$.

\section{Yang-Mills on $S^1 \times S^1 \times S^2$ with 
non-zero chromomagnetic field on $\RR^2$}
Superficially, it would seem that there are
two different configurations of chromomagnetic field that
can exist on the manifold $S^1 \times S^1 \times S^2$-- either with the 
chromomagnetic field polarized along the torus $S^1 \times S^1$ or
along the sphere $S^2$. The first case cannot be realized because
a covariantly constant chromomagnetic field can not exist on 
$S^1 \times S^1$.
However, we can consider the related
problem of having the non-zero field polarized along $\RR^2$ on the
manifold $\RR^2 \times S^2$. This case is not physical because it
leaves the time direction to be incorporated into $S^2$, which 
means that spacetime can no longer be deforemed to have the 
structure $\RR \times \Sigma$.
This case has been investigated by Elizalde, et. al. \cite{elizalde96b}.
In the limit that the curvature is small, it has been determined that
the vacuum stabilizes for some radius of $S^2$. We can also analyze this
problem from our standpoint.

In the case of a chromomagnetic field directed along $\RR^2$,
the operator $L_{YM}$ will have the block diagonal form
\be
L_{YM} = (-\Delta_{1(1)} - \Delta_{1(2)})
\left(
\begin{array}{cc}
\II & 0 \\
0 & \II
\end{array}
\right)
+ \left(
\begin{array}{cc}
0 & 0 \\
0 & \R_2
\end{array}
\right)
- 2\left(
\begin{array}{cc}
\F_1 & 0 \\
0 & 0
\end{array}
\right)\quad,
\ee
where $\Delta_{1(1)}$ is the Laplacian acting on one-forms on 
$\RR^2$, $\Delta_{1(2)}$ is the Laplacian acting on 
one-forms on $S^2$, $\R_2$ is the Ricci tensor on $S^2$, and 
$\F_1$ is the chromomagnetic field restricted to 
$\RR^2$.
The operator $L_{FP}$ is simply the Laplacian acting on scalars
\be
L_{FP} = -\Delta_{0(1)} - \Delta_{0(2)}\,,
\ee
where $\Delta_{0(1)}$ is the Laplacian acting on scalars on 
$\RR^2$ and $\Delta_{1(2)}$ is the Laplacian acting on 
scalars on $S^2$.

The total heat kernel $U_{\rm tot}(t)$ is then 
\bea
U_{\rm tot}(t) &=& \Tr \exp(-tL_{YM}){}_{S^2}\times\Tr \exp(-tL_{YM}){}_{\RR^2}
\nonumber\\
&&
-2\Tr \exp(-tL_{FP}){}_{S^2}\times\Tr \exp(-tL_{FP}){}_{\RR^2}
\eea
Using the heat trace expressions (\ref{R2_YM}), (\ref{R2_FP}), (\ref{S2_H0}), (\ref{S2_H0g}), this gives 
the result
\bea
U_{\rm tot}(t) &=& \int_{\RR^2}\,dx (4\pi t)^{-1} \Bigg\{
\left[2\Theta_1\left(\frac{t}{R^{2}}\right) + \Theta_0\left(\frac{t}{R^{2}}\right)\right]
\nonumber\\
&&
\times \left(\frac{tHR^{-2}}{\sinh(tHR^{-2}/2)}\left\{4e^{-t/R^{2}}
\left[2+4\sinh^2\left(\frac{tHR^{-2}}{2}\right)\right] -4 \right\} + 8e^{-t/R^2} - 4\right)
\nonumber\\
&&
-4\left[1+\frac{tHR^{-2}}{\sinh(tHR^{-2}/2)}\right]
\Bigg\}
\eea
\section{Yang-Mills on $S^1 \times S^1 \times S^2$ with 
non-zero chromomagnetic field on $S^2$}
Another allowable configuration is to let the chromomagnetic field 
lie along $S^2$. In this case, the total heat kernel is given by
\bea
&&
U_{\rm tot}(t) = \Tr \exp(-tL_{YM}){}_{S^2}\times\Tr \exp(-tL_{YM}){}_{S^1 \times S^1}
\nonumber\\
&&\qquad
-2\Tr \exp(-tL_{FP}){}_{S^2}\times\Tr \exp(-tL_{FP}){}_{S^1 \times S^1}
\eea
Using the $T^2$ heat trace expressions (\ref{T2_YM}),(\ref{T2_FP}) with H = 0, and the $S^2$
heat trace expressions 
(\ref{S2_H0})-(\ref{S2_H6}) calculated in Chapter 4, this gives us 
the following results:\\
For $H=0$,
\bea
U_{\rm tot}(t) &=& 8S\left(\frac{t}{r_1^2}\right)\,S\left(\frac{t}{r_2^2}\right)
\nonumber\\
&&
\times \left\{\left[2e^{-t/R^2} - 1\right]
\left[2\Theta_1\left(\frac{t}{R^2}\right) + \Theta_0\left(\frac{t}{R^2}\right)\right] -1\right\}\,.
\eea
For $H=1$,
\bea
U_{\rm tot}(t) &=&  8S\left(\frac{t}{r_1^2}\right)S\left(\frac{t}{r_2^2}\right)
\nonumber\\
&&
\times \Bigg\{3e^{-(9/4)t/R^2}\Phi_0\left(\frac{t}{R^2}\right)
+ (4\cosh(t/R^2)e^{-t/R^2} -2)\Theta_1\left(\frac{t}{R^2}\right)
\nonumber\\
&&
-2e^{-2t/R^2}\Theta_0\left(\frac{t}{R^2}\right) -6e^{-9t/4R^2}
\Bigg\}\,.
\nonumber\\
&&
\eea
For $H=2$,
\bea
U_{\rm tot}(t) &=&  4S\left(\frac{t}{r_1^2}\right)S\left(\frac{t}{r_2^2}\right)
\nonumber\\
&&
\times
\Bigg\{
\left(4e^{-t/R^2}\cosh(2t/R^2) - 4\right)
\left[2\Theta_1\left(\frac{t}{R^2}\right) + \Theta_0\left(\frac{t}{R^2}\right)\right]
\nonumber\\
&&
+2e^{t/R^2} - 6e^{-3t/R^2}
\Bigg\}\,.
\eea
For $H=3$,
\bea
&&
U_{\rm tot}(t) =  4S\left(\frac{t}{r_1^2}\right)S\left(\frac{t}{r_2^2}\right)
\nonumber\\
&&
\qquad
\times
\Bigg[
(2e^{7t/4R^2}+10e^{-19t/4R^2} -6e^{t/4R^2})\Phi_0\left(\frac{t}{R^2}\right)
\nonumber\\
&&
\qquad
\left(8\cosh(3t/R^2)e^{-t/R^2} -4\right)\Theta_1\left(\frac{t}{R^2}\right) + 4\Theta_0\left(\frac{t}{R^2}\right)
\nonumber\\
&&
-20e^{-8t/R^2} -20e^{-11t/R^2} -8e^{-6t/R^2} -8e^{-10t/R^2}) + 6e^{-3t/4R^2}
\Bigg]
\eea
For $H=4$,
\bea
U_{\rm tot}(t) &=&  4S\left(\frac{t}{r_1^2}\right)S\left(\frac{t}{r_2^2}\right)
\nonumber\\
&&
\times
\Bigg\{
\left[4e^{-t/R^2}\cosh(4t/R^2) - 2\right]
\left[4\Theta_1\left(\frac{t}{R^2}\right) + \Theta_0\left(\frac{t}{R^2}\right)\right]
\nonumber\\
&&
-6e^{-7t/R^2} +10e^{-11t/R^2} + 6e^{-2t/R^2}
\Bigg\}\,.
\eea
For $H$ odd, $H\ge5$,
\bea
U_{\rm tot}(t) &=&  4S\left(\frac{t}{r_1^2}\right)\,S\left(\frac{t}{r_2^2}\right)
\nonumber\\
&&
\times
\Bigg\{
(8\cosh(tH/R^2)e^{-t/R^2}-4)\Theta_1\left(\frac{t}{R^2}\right)
\nonumber\\
&&
+ \left[(2H-2)-(4H+4)\cosh(tH/R^2)e^{-t/R^2}\right]\Theta_0\left(\frac{t}{R^2}\right)
\nonumber\\
&&
\left[(4H+8)\cosh(tH/R^2)e^{-t/R^2}-8e^{-tH/R^2}e^{-t/R^2}-2H\right]e^{t/4R^2}\Phi_0\left(\frac{t}{R^2}\right)
\nonumber\\
&&
-2e^{-t(H+3/4)/R^2}(H+2)\sum_{l=1}^{\frac{H}{2}+\frac{1}{2}}e^{-tl^2/R^2}
\nonumber\\
&&
-2e^{t(H-3/4)/R^2}(H-2) \sum_{l=1}^{\frac{H}{2}-\frac{3}{2}}e^{-tl^2/R^2}
\nonumber\\
&&
-2e^{-t(H+1)/R^2}\sum_{l=1}^{\frac{H}{2}+\frac{1}{2}}(2l+H-1)e^{-tl(l+1)/R^2}
\nonumber\\
&&
-2e^{t(H-1)/R^2} \sum_{l=1}^{\frac{H}{2}-\frac{3}{2}}(2l+3-H)e^{-tl(l+1)/R^2}
\nonumber\\
&&
+2He^{t/4R^2}\sum_{l=1}^{\frac{H}{2}-\frac{1}{2}}e^{-tl^2/R^2}
+\sum_{l=1}^{\frac{H}{2}-\frac{1}{2}}(4l+2-2H)e^{-tl(l+1)/R^2}
\Bigg\}
\eea
For $H$ even, $H\ge6$,
\bea
U_{\rm tot}(t) &=&  4S\left(\frac{t}{r_1^2}\right)S\left(\frac{t}{r_2^2}\right)
\nonumber\\
&&
\times
\Bigg\{
\left(4e^{-t/R^2}\cosh(tH/R^2) - 2\right)\left[2\Theta_1\left(\frac{t}{R^2}\right) - \Theta_0\left(\frac{t}{R^2}\right)\right]
\nonumber\\
&&
-2e^{t(H-1)/R^2}\sum_{l=1}^{\frac{H}{2}-2}(2l+1)e^{-tl(l+1)}
-2e^{-t(H+1)/R^2}\sum_{l=1}^{\frac{H}{2}}(2l+1)e^{-tl(l+1)}
\nonumber\\
&&
+2\sum_{l=1}^{\frac{H}{2}-1}(2l+1)e^{-tl(l+1)}
\Bigg\}
\eea
\section{Stability}
When the heat traces above contain an exponential that 
grows or stays constant with $t$,
then the Yang-Mills vacuum will be unstable, causing the 
configuration with constant chromomagnetic field to
decay into another state. However,
if all exponentials are decreasing, then the vacuum will be 
stable. 


If the chromomagnetic field is polarized along $S^2$ on the manifold,
the stability of the the constant chromomagnetic state will depend on the strength $H$ 
of the chromomagnetic field and the radius $R$ of the sphere $S^2$.
The case $H=1$ will always be stable, but $H=2$, and $H=3$ will not 
be stable for any radii. In the case of $H=1$, 
the lowest eigenvalue is given by 
\be
\lambda_{\rm min} = \frac{1}{R^2}\left(\frac{5}{4}\right)\,,
\ee
which implies that the vacuum stabilizes for $H=1$. 
Similarly, for $H=2$, the minimum eigenvalue is
\be
\lambda_{\rm min} = -\frac{1}{R^2}\,,
\ee
and the vacuum is unstable. 
For $H=3$, the minimum eigenvalue is 
\be
\lambda_{\rm min} = -\frac{3}{4R^2}\,
\ee
and the vacuum is unstable.
For $H\ge4$, the lowest mode will correspond to the eigenvalue
\be
\lambda_{\rm min} = \frac{1}{R^2}\left[\left(\frac{H}{2} - 1\right)\frac{H}{2} - H + 1\right]\,.
\ee
The vacuum will be stable when this eigenvalue is positive, which 
occurs when the condition 
\be
\frac{H^2}{4}-\frac{3H}{2} + 1\ge 0
\ee
is satisfied. This occurs for 
\be
H \ge 6\,.
\ee
Thus, the vacuum is unstable for $H = 2,3,4,5$ and stable for
$H=0,1$ and $H \ge 6$.
Because $H$ is the dimensionless parameter relating to the magnetic
field $M$, $H = MR^2$, this implies that we can make a configuration
stable by increasing either the magnetic field or the radius. A large 
radius would intuitively return us to the Saviddy flat-space case, but 
we instead see that it actually increases the lowest eigenvalue. 
The local behavior of these cases is the same, so this must
be a topological phenomenon.

\section{Effective Action}
If we consider the radii of the spheres to be variable,
then the lowest energy state is 
state that minimizes the effective action, which will
be a function of both the chromomagnetic field and 
the radius of the spheres. The one-loop effective action is written
in terms of the heat kernel as 
\be
\Gamma_{(1)} = -\frac{1}{2}\frac{d}{dp}\left[\frac{\mu^{2p}}{\Gamma(p)}\int_0^\infty dt\,t^{p-1}U_{\rm tot}(t)\right]_{p=0}
\ee
The effective action, then, to first order is 
\be
\Gamma = S + \hbar \Gamma_{(1)}\,
\ee
where $S$ is the classical action
\be
S = -\frac{1}{8e^2}\int_M dx\, \tr{\F^{\mu\nu}\F_{\mu\nu}}\,,  
\ee
which in our case can be integrated over the manifold to give
\be
S = \frac{H^2}{8e^2R^2}\vol(M)\,.
\ee
Calculating the effective action and finding a global minimum for
all $R$ and $H$ will reveal the vacuum with minimum energy.
\chapter{CONCLUSION}

We have calculated the heat traces for pure Yang-Mills on 
products of spheres with a covariantly constant
chromomagnetic field, and have shown that for a space 
with certain values of curvature and magnetic field, the covariantly 
constant chromomagnetic vacuum forms a local minimum of the
effective action. This lends creedence to the possibility 
that the Savvidy-type vacuum, with a covariantly constant
magnetic field will form an absolute minimum on the 
relevant spaces. 

Contrary to expectations, the limit in which the radius 
of the sphere becomes infinite does not yield the standard
flat-space results for eigenvalues of the Laplacian.
Instead of having eigenvalues that are linear in the 
magnetic field, our results show that on the two-sphere,
the lowest eigenvalue of the Laplacian will increase 
quadratically with the magnetic field and quadratically
with the radius of the sphere, leading to the 
counter-intuitive result that the space most closely
approximating the flat-space Saviddy vacuum will have
a minimum eigenvalue that will be farthest from being
unstable. This effect is a topological phenomenon that
requires further study. 

The next step in examining this model should be
to calculate the full effective potential as a function of 
both the strength of the chromomagnetic field and the 
radius of the sphere. The absolute minimum of the effective
potential would yield the absolute vacuum state of Yang-Mills. 
In our case, we would find a state that would be at 
least a local minimum of the vacuum, and possibly the absolute
minimum.

\appendix
\chapter{EIGENVALUES AND DEGENERACIES OF $-\Delta$}
In this section, we find the eigenfunctions and the
corresponding eigenvalues $\lambda_l$ of the operator
$-\Delta$. They are given by regular solutions of the
equation
\be
\left\{-\frac{1}{R^2\sin\theta}\left[\sin\theta\partial_\theta^2
+ \cos\theta \partial_\theta + \frac{1}{\sin\theta}
(i m - i k_j\cos\theta)^2\right] - \lambda_l \right\}u(\theta) = 0\,.
\ee
The label $l$ on $\lambda_l$ is only a label here. It's allowed
values will be found later.
Introducing the change of variables
$x = \cos\theta$, the equation becomes
\be
\left[\partial_x (1-x^2) \partial_x -
        \frac{1}{1-x^2}(m - k x)^2
        + R^2\lambda\right]u(x) = 0\,.
\ee
We may make the substitution
\be
u(x) = (1-x)^{\alpha}(1+x)^{\beta}f(x)\,,
\ee
where
\be
\alpha = \left|\frac{m-k}{2}\right|\,, \quad
\beta = \left|\frac{m+k}{2}\right|\,.
\ee
to get the equation
\bea
&&
\Biggl\{(1-x^2)\frac{d^2}{dx^2} - [(2\beta - 2\alpha)
+ (-2-2\alpha -2\beta)x]\frac{d}{dx}
\nonumber\\
&&
\qquad\qquad\qquad
+ \left[-\alpha -\beta -(\alpha + \beta)^2
+ R^2\lambda\right]\Biggr\} f(x) = 0\,.
\eea
By switching variables to $z = \frac{1-x}{2}$, we obtain
the hypergeometric equation
\bea
&&
\Biggl\{z(1-z)\frac{d^2}{dz^2} + [(1+2\alpha) - 2(1 + \alpha + \beta)z]\frac{d}{dz}
\nonumber\\
&&
\qquad\qquad\qquad
+ [-\alpha -\beta -(\alpha + \beta)^2 + R^2\lambda]\Biggr\} f(z) = 0\,.
\eea
Finally, introducing the notation
\be
a^\pm = \frac{1}{2} + \alpha + \beta
    \pm \frac{1}{2}(1 + 4R^2\lambda)^{1/2}\,,
\ee
we obtain the solution
\be
f(z) = F\left(a^+,a^-; 1+\left|m-k\right|;\, z\right)\,,
\ee
where the function $F$ is the hypergeometric
function \cite{erdelyi53}
\be
\label{hypergeofn}
F(a,b;c;z) = \sum_{n=0}^\infty \frac{(a)_n(b)_n}{n!(c)_n}z^n\,.
\ee
The notation $(a)_n$ denotes the Pochhammer symbol
\be
(a)_n = \frac{\Gamma(a+n)}{\Gamma(a)}\,.
\ee
In our case, it is easy to show that $c > 0$ and
$a+b-c \ge 0$, so the expression (\ref{hypergeofn})
will diverge at $|z| = 1$ unless the series terminates \cite{erdelyi53}.
Regular solutions will exist only in the degenerate case when
the series terminates and the hypergeometric
function becomes a polynomial. This happens when at
least one of the first two arguments of $F$ is a
negative integer.
Thus, the only regular solutions occur when
\be
a^- = \frac{1}{2} + \alpha + \beta
    - \frac{1}{2}(1 + 4R^2\lambda)^{1/2} = -q\,,
    \quad q = 0,1,2,\dots
\ee
where $q$ is a non-negative integer.
This yields the eigenvalues
\be
\label{lambda}
\lambda = \frac{1}{R^2}(\alpha + \beta + q)
    (\alpha + \beta + q + 1)\,.
\ee
The quantity $\alpha + \beta$ takes on the values
\be
\alpha + \beta = \frac{1}{2}\left\{|m+k| + |m-k|\right\} =
\max\left\{|m|,|k|\right\}\,.
\ee
Then it can be directly seen from (\ref{lambda})
that the lowest eigenvalue $\lambda_0$ is
\be
\lambda_0 = \frac{1}{R^2}|k|(|k|+1)\,.
\ee

For integer values of $k$, the quantity $\alpha + \beta$
is always integer, and so the eigenvalues are
\be
\label{eval_int}
\lambda_l = \frac{1}{R^2}(|k|+l)(|k|+l+1)\,,\qquad l = 0,1,2,\dots
\ee
Degeneracies of these eigenvalues can be counted using
(\ref{lambda}), (\ref{eval_int}), and the fact that $m$ is
an integer. For any given value
of $l$, the $2|k| + 1$ cases $|m|\le|k|$ correspond to $q = l$.
The $2l$ cases $m = \pm(|k|+l),\pm(|k|+l-1),\dots,\pm(|k|+1)$
correspond to $q = 0,1,\dots,l-1$, respectively.
Counting these cases, the total degeneracy $d_l$ of
$\lambda_l$ from (\ref{eval_int}) is
\be
d_l = 2(|k|+l) + 1\,.
\ee

Now consider the case where $k$ is a half-integer.
Then there are two cases: when the quantity
$\alpha + \beta$ is a half-integer, and when
$\alpha + \beta$ is an integer.

First consider the case when $\alpha + \beta$
is half-integer. The eigenvalues are then
\be
\lambda_l = \frac{1}{R^2}(|k|+l)(|k|+l+1)\,,\qquad l = 0,1,2,\dots
\ee
In this case, we must have
$\alpha + \beta = |k|$, which corresponds to the cases
$|m| \le |k|$. The number of integer values of $m$ that satisfy
this inequality is
\be
d_l = 2|k|\,.
\ee

In the second case,
$\alpha + \beta $ is integer, in which case
$\alpha + \beta = |m|$. Because $\alpha + \beta$ and
$q$ are integers, the eigenvalues written in terms of $k$ are
\be
\lambda_l = \frac{1}{R^2}\left(|k|+l + \frac{1}{2}\right)
\left(|k|+l+\frac{3}{2}\right)\,,\qquad l = 0,1,2,\dots
\ee
The $2l + 2$ cases
\be
m = \pm\left(|k|+\frac{1}{2}\right),
\pm\left(|k|+\frac{1}{2} + 1\right),\dots,
\pm\left(|k|+\frac{1}{2} + 2\right)
\ee
correspond to $q= l,l-1,\dots,0$, respectively.
Counting these gives the degeneracies
\be
d_l = 2l+2\,.
\ee

\begin{Bibliographyno}








\bibitem{avramidi93}
I. G. Avramidi, {\it  A new algebraic approach for calculating the heat
kernel in gauge theories}, Phys. Lett. B {\bf 305} (1993) 27--34.

\bibitem{avramidi94}
I. G. Avramidi,  {\it The heat kernel on symmetric spaces via
integrating over the group  of isometries}, Phys. Lett. B {\bf 336}
(1994) 171--177.

\bibitem{avramidi95a}
I. G. Avramidi, {\it Covariant algebraic calculation of the one-loop
effective  potential in non-Abelian gauge theory and a new approach to
stability problem}, J. Math. Phys. {\bf 36} (1995) 1557--1571.

\bibitem{avramidi99}
I. G. Avramidi, {\it A model of stable chromomagnetic vacuum in
higher-dimensional Yang-Mills theory},  Fortschr. Phys.,  {\bf 47}
(1999) 433--455.

\bibitem{avramidi00}
I. G. Avramidi, {\it Heat Kernel and Quantum Gravity},
Lecture Notes in Physics, Series Monographs, LNP: m64 (Berlin:
Springer-Verlag,  2000)

\bibitem{avramidi02}
I. G. Avramidi, {\it Heat kernel in quantum field theory},
Nucl. Phys. B - Proc. Suppl. {\bf 104} (2002) 3-32

\bibitem{dewitt65}
B. S. De Witt, {\it Dynamical Theory of Groups and Fields}, (New York: Gordon and
Breach, 1965).

\bibitem{dewitt03}
B. S. DeWitt, {\it The Global Approach to Quantum Field Theory},
(Oxford: Oxford University Press, 2003).

\bibitem{elizalde96b}
Emilio Elizalde, S. Odintsov, and A. Romeo, 
{\it Effective potential for a covariantly constant gauge
field in curved spacetime} arXiv:hep-th/9607189

\bibitem{erdelyi53}
A. Erd\'elyi, W. Magnus, F. Oberhettinger and F. G. Tricomi,  {\it
Higher Transcendental Functions}, (New York: McGraw-Hill, 1953), vol. I.

\bibitem{frankel97}
T. Frankel, {\it The Geometry of Physics},
(Cambridge: Cambridge University Press, 1997).

\bibitem{gross73}
D. Gross and F. Wilczek,
{\it Ultraviolet Behavior of Non-Abelian Gauge Theories},
Phys. Rev. Lett. {\bf 30} (1973) 1343--1346.

\bibitem{landau77}
L. D. Landau and E. M. Lifschitz, {\it Quantum Mechanics},
(Oxford: Butterworth Heinemann, 1977).

\bibitem{nielsen79b}
H.B. Nielsen and M. Ninomiya,
{\it A Bound on Bag Constant and Nielsen-Olesen Unstable Mode in QCD},
Nucl. Phys. {\bf B156} (1979) 1--28.

\bibitem{nielsen79}
H. B. Nielsen and P. Olesen, {\it A quantum liquid model for the QCD
vacuum: gauge and rotational invariance of domained and quantized
homogeneous color fields}, Nucl. Phys. {\bf B160} (1979) 380--396.

\bibitem{nielsen78}
N. K. Nielsen and P. Olesen, {\it  An unstable Yang-Mills mode}, Nucl.
Phys. {\bf B 144} (1978) 376--396.

\bibitem{politzer73}
H. D. Politzer, {\it Reliable Perturbative Results for Strong Interactions}, Phys. Rev.
Lett. {\bf 30} (1973) 1346--1349.

\bibitem{savvidy77}
G. K. Savvidy, {\it Infrared instability of the
vacuum state of gauge theories and asymptotic freedom}, Phys. Lett.
{\bf B 71} (1977) 133--134.

\bibitem{vassilevich03}
    D.V. Vassilevich,
    Heat Kernel Expansion: User's Manual,
    \textit{Phys. Rep.}, \textbf{388}~(2003), no. 5--6, 279--360,
    arXiv:hep-th/0306138.





\bibitem{vilkovisky_gospel}
G. A. Vilkovisky, {\it The Gospel according to De~Witt}, in: {\it
Quantum Gravity}, Ed. S. Christensen, (Bristol: Hilger, 1983), pp.
169--209.

\end{Bibliographyno}
\end{document}